\documentclass[aps,prb,twocolumn,showpacs]{revtex4}  
\usepackage{amsmath,amssymb,amstext,graphicx,bm}
\makeatletter

\begin{document}
\title{Electric Dipole Induced Spin Resonance in Quantum Dots}
\author{Vitaly N. Golovach}
\author{Massoud Borhani}
\author{Daniel Loss}
\affiliation{Department of Physics and Astronomy, 
University of Basel, Klingelbergstrasse 82, 
4056 Basel, Switzerland}

\date{\today}

\begin{abstract}
An alternating electric field, applied to a quantum dot, 
couples to the electron spin via the spin-orbit interaction.
We analyze different types of spin-orbit coupling known in the literature and 
find two efficient mechanisms of spin control in quantum dots.
The linear in momentum Dresselhaus and Rashba spin-orbit couplings give rise to
a fully transverse effective magnetic field in the presence of a Zeeman splitting
at lowest order in the spin-orbit interaction.
The cubic in momentum Dresselhaus terms are efficient 
in a quantum dot with non-harmonic confining potential and give rise to a spin-electric 
coupling proportional to the orbital magnetic field.
We derive an effective spin Hamiltonian, which can be used to implement
spin manipulation on a timescale of $10\,{\rm ns}$ with the current experimental setups.
\end{abstract}
\maketitle

\section{Introduction}
Coherent manipulation of electron spin is at the heart of 
spintronics~\cite{Wolf,ALS} and
quantum computing with spins.~\cite{LD}
In the proposal of Ref.~\onlinecite{LD}, the spin of an electron
confined to a quantum dot is used as qubit to store and process quantum information.
A quantum register consisting of an array of such spin-$1/2$ quantum dots is operated
by a set of quantum gates that act on single spins and pairs of neighboring spins.~\cite{LD}
Among the simplest quantum gates are the spin rotations on the Bloch sphere.
With the help of only a static magnetic field and an electron-spin-resonance (ESR) pulse, 
one can change the state of the spin qubit at will.
It is important, however, that the ESR pulse can be applied locally to each of the quantum dots,
ensuring that the spins are accessed independently from one another.
For an ESR~\cite{Atherton,Engel} to occur, usually, the electron is exposed to 
an alternating magnetic field of a frequency $\omega_{\it ac}$ that matches 
the electron Zeeman splitting.
However, because strong local electric fields are easier to obtain than strong local magnetic fields,
interest arises in spin resonance induced by electric fields.

Recently, Kato~{\em et al.}~\cite{Kato} have demonstrated three-dimensional control 
of spins in a GaAs/Al${}_x$Ga${}_{1-x}$As heterostructure with the use of an 
alternating electric field.
The mechanism of spin coupling to the electric field relies on a specially
engineered Land\'e $g$ tensor in the heterostructure, achieved by modulating
the Al content during the MBE growth.~\cite{Miller}
The resulting $g$ tensor is both anisotropic and space-dependent, and allows
control over the direction and magnitude of the spin precession frequency.~\cite{Kato,Salis}
A $g$-factor modulation resonance ($g$-TMR) occurs 
similarly to an ESR, when the frequency of the electric field matches 
the Zeeman splitting.~\cite{Kato}  
Rashba and Efros~\cite{RashbaEfros} have further proposed to use 
the standard (Dresselhaus~\cite{Dress} and Rashba~\cite{Rashba}) spin-orbit couplings to achieve an 
electric dipole induced spin resonance (EDSR) in quantum wells.
Rashba and Efros~\cite{RashbaEfros} have shown that the EDSR
is highly efficient in quantum wells, promising electron spin control on a timescale
$\omega_R^{-1}\simeq 100\,{\rm ps}$, where $\omega_R$ is the Rabi frequency.~\cite{Atherton}
These results have important practical implications in spintronics, where spins of extended electrons
are used as a resource to accomplish information processing.
In the context of quantum computing, however, interest arises in the spin of a localized electron.
A natural question is, therefore, 
``What is the microscopic mechanism of EDSR in quantum dots and how strong is the EDSR effect?''

EDSR has nearly half-a-century long history. 
It has been first observed for extended electrons in bulk semiconductors,~\cite{Bell,McCombe} and
studied more recently for donor-bound electrons in Cd${}_{1-x}$Mn${}_x$Se~\cite{Dobrowolska} and extended
electrons in two-dimensional electron gases~\cite{Schulte,DuckheimLossEDSR} and epilayers.~\cite{KatoEDSR}
The ``forbidden'' electric-dipole transition between the electron spin-up and spin-down states
becomes possible in the presence of spin-orbit interaction.
Absorption spectra of EDSR provide information about the value of the electron $g$ factor and
the strength of the spin-orbit coupling.
In two-dimensional electron systems, one expects the Dresselhaus spin-orbit 
interaction~\cite{Dress} to be enhanced compared to
bulk semiconductors, because of the confinement of electron motion in one direction. 
Furthermore, the Rashba spin-orbit interaction~\cite{Rashba} arises in heterostructures
lacking inversion symmetry, such as, e.g., heterojunctions.
In some systems, the Rashba coupling constant can be efficiently tuned by electric fields.~\cite{Nitta}

In quantum dots,~\cite{KAT} the spin-orbit interaction is generally suppressed
due to complete localization of electron motion.~\cite{KhNaHaAlFa}
Typically, the quantum dot lateral size $\lambda_d$ is smaller than the spin-orbit length $\lambda_{SO}$,
and any effect of the spin-orbit interaction is suppressed as a power of $\lambda_d/\lambda_{SO}$
and therefore is expected to be weak.
This expectation contrasts with the case of electrons in quantum wells, where
the EDSR meets most favorable conditions.~\cite{RashbaEfros}
The Zeeman interaction in quantum dots plays an important role for observing spin-orbit effects.~\cite{GKL,KhaNaz}
Without the Zeeman interaction, the Rashba and linear in momentum Dresselhaus spin-orbit terms 
do not contribute to spin-related phenomena at the first order of spin-orbit interaction.
This ``absence of spin-orbit'' at the leading order in quantum dots has been discussed extensively
in the literature.~\cite{KhNaHaAlFa,GKL,KhaNaz}
Below, we show that a similar result arises also for the cubic in momentum Dresselhaus terms in the case when the dot
confining potential is quadratic and the perturbation is linear in the electron coordinates.

In this paper, we consider the use of EDSR for control of individual electron spins in quantum dots.
We derive an effective spin Hamiltonian for a quantum-dot electron, subject to {\em ac} electric fields.
We show that there are two major mechanisms of EDSR in quantum dots.
One arises from the linear in momentum Dresselhaus and Rashba spin-orbit couplings in combination with
the Zeeman interaction.
The other arises from the cubic Dresselhaus terms in combination with the cyclotron frequency.
We estimate the strengths of both EDSR effects and compare them to the ordinary ESR.
We find that despite a strong suppression, compared to quantum wells, the
EDSR in quantum dots is still an efficient mechanism of spin manipulation and can be used
alone or together with ESR to achieve control of spin on a timescale $\omega_R^{-1}\simeq 10\,{\rm ns}$.

\begin{figure}
\begin{center}
\includegraphics[angle=0,width=0.45\textwidth]{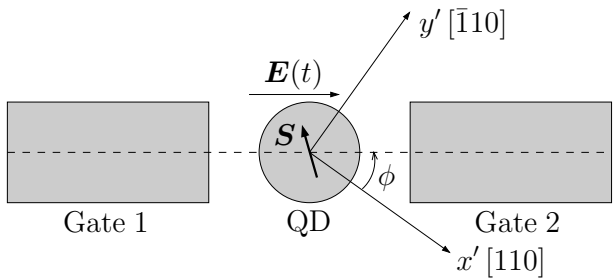}
\caption{\small
Schematic of a setup for electric field control of spin via the spin-orbit interaction.
The quantum dot (QD) contains a single electron with spin
$\mbox{\boldmath $S$}=(\hbar/2)\mbox{\boldmath $\sigma$}$, deep in the Coulomb blockade valley.
The gates $1$ and $2$ are used to generate an alternating
electric field $\mbox{\boldmath $E$}(t)$, which acts via the 
spin-orbit interaction on the electron spin.
As a result, an electric dipole spin resonance (EDSR)
occurs if the frequency of $\mbox{\boldmath $E$}(t)$
is tuned to match the Larmor frequency $\omega_Z=E_Z/\hbar$.
}
\label{EDSRfigure}
\end{center}
\end{figure}

\section{EDSR setup}
We consider a quantum dot containing a single electron with charge $-e$ and spin 
$\mbox{\boldmath $S$}=(\hbar/ 2)\mbox{\boldmath $\sigma$}$, where 
$\mbox{\boldmath $\sigma$}=(\sigma_x,\sigma_y,\sigma_z)$ are the Pauli matrices.
The quantum dot is in the Coulomb blockade regime with
extraction ($U_-$) and addition ($U_+$) energies that are large
compared to the temperature, so that the dot occupation remains constant. 
An external electric field $\mbox{\boldmath $E$}(\mbox{\boldmath $r$},t)$ is applied
to the quantum dot. In practice, $\mbox{\boldmath $E$}(\mbox{\boldmath $r$},t)$ 
can be generated by a pair of gates, as sketched in Fig.~\ref{EDSRfigure}, 
to which an {\em ac} signal of frequency $\omega_{\it ac}$ is supplied from an external circuit (not shown).
The Hamiltonian describing the quantum dot electron in the external alternating field reads
\begin{equation}
H=H_0+V(\mbox{\boldmath $r$},t),
\label{HamH}
\end{equation}
where 
$V(\mbox{\boldmath $r$},t)=e\int^{\bm r} d\mbox{\boldmath $r$}'\cdot\mbox{\boldmath $E$}(\mbox{\boldmath $r$}',t)$ 
is the potential energy of the electron in the external electric field 
and $H_0$ is the ``unperturbed'' Hamiltonian (see further).
In particular, for an electric field constant in space 
$\mbox{\boldmath $E$}(\mbox{\boldmath $r$},t)=\mbox{\boldmath $E$}(t)$, 
the potential energy reads $V(\mbox{\boldmath $r$},t)=e\mbox{\boldmath $E$}(t)\cdot\mbox{\boldmath $r$}$.

For practical applications, it is a good idea to use two gates,
as shown in Fig.~\ref{EDSRfigure}, because this allows larger amplitudes of $\mbox{\boldmath $E$}(t)$ to be applied
to the quantum dot, while still maintaining the dot within the same Coulomb blockade valley.
Ideally, the {\em ac} voltage drop is distributed between the two gates symmetrically
and the dot potential is kept constant by counteracting potential shifts quadratic in the electric field.
For a harmonic quantum dot, the desired {\em ac} potential reads
\begin{equation}
V(\mbox{\boldmath $r$},t)=e\mbox{\boldmath $E$}(t)\cdot\mbox{\boldmath $r$}+\frac{\left(e E(t)\right)^2}{2m_e\omega_0^2},
\end{equation}
where $m_e$ is the electron effective mass and $\omega_0$ is the oscillator frequency.
Then the only effect of the {\em ac} signal on the dot confinement is shifting the dot center
as a function of time by the amount
\begin{equation}
\mbox{\boldmath $r$}_0(t)=-\frac{e\mbox{\boldmath $E$}(t)}{m_e\omega_0^2}.
\label{eqr0}
\end{equation} 
The amplitude of $\mbox{\boldmath $r$}_0(t)$ is going to be a relevant parameter in
our following analysis. 
Therefore, setups in which the dot can be easily moved on the substrate by gates 
are particularly interesting in the context of this paper.
We discuss the case of $r_0\sim\lambda_{\it SO}$ in Sec.~\ref{secDisc},
whereas for the bulk of the paper we restrict ourself to $r_0\ll\lambda_{\it SO}$.

The Hamiltonian $H_0$ consists of several terms,
\begin{equation}
H_0=H_d+H_Z+H_{SO},\label{HamH0}
\end{equation}
where $H_d$ is the Hamiltonian of a confined electron,
\begin{equation}
H_d=\frac{ p^2}{2m_e}+U(\bm{r}),\label{Hdot}
\end{equation}
with ${\bm p} = -i \hbar \partial/\partial{\bm r} + (e/c)\bm{A}(\bm{r})$
being the electron momentum, $c$ the speed of light in vacuum, 
and $U(\mbox{\boldmath $r$})$ the quantum dot confining potential.
We restrict our consideration to quantum dots with strong
confinement along one axis, such as, e.g., quantum dots defined in
a two-dimensional electron gas (2DEG).
For GaAs, the 2DEG lies, typically, in the crystallographic plane $(001)$
and has a width $d\simeq 5\,{\rm nm}$, which ensures a strong size quantization
along $z\parallel[001]$.
The in-plane motion of the electron is described by the Hamiltonian (\ref{Hdot}),
where ${\bm r} = (x,y)$ is the electron in-plane coordinate; whereas the transverse
motion (along $z$) has already been integrated out in Eqs.~(\ref{HamH})-(\ref{Hdot}).
In the absence of external time-dependent fields,
${\bm A}({\bm r})$ accounts for the orbital effect of a static magnetic field $\mbox{\boldmath $B$}$. 
Assuming that ${\bm B}$ is constant in space,
we have ${\bm A}({\bm r})=B_z(-y/2,x/2,0)$ in the symmetric gauge.
Note that the in-plane components $B_x$ and $B_y$ are not present in ${\bm A}({\bm r})$,
because the motion along $z$ is strongly quantized ($d\ll\sqrt{\hbar c/e B_{x(y)}}$).

The magnetic field ${\bm B}$ also induces a Zeeman splitting $E_Z=g\mu_BB$
and a spin quantization axis ${\bm n}={\bm B}/B$ via the Zeeman interaction,
\begin{eqnarray}
H_Z&=&\frac{1}{2}g\mu_B{{\bm B}}\cdot\mbox{\boldmath $\sigma$}
=\frac{1}{2}E_Z{{\bm n}}\cdot\mbox{\boldmath $\sigma$},
\label{HZ}
\end{eqnarray}
where $g$ is the electron $g$-factor and $\mu_B$ is the Bohr magneton.
In GaAs, the magnitude of the $g$-factor is anomalously small ($g\approx -0.44$) compared
to other ${\rm A}_{\mbox{\scriptsize III}}{\rm B}_{\mbox{\scriptsize V}}$ semiconductors.
The Zeeman energy is, therefore, 
much smaller than the cyclotron energy $\hbar\omega_c=\hbar eB_z/m_ec$ by a factor $gm_e/m\ll 1$
(with $m$ being the electron mass in vacuum), 
for magnetic fields applied transversely to the 2DEG. 
In Sec.~\ref{secSOp3}, we derive an efficient spin-electric coupling that is
proportional to $\hbar\omega_c$, but present only in non-harmonic quantum dots.
We remark that the magnetic field is an important ingredient in our
EDSR scheme, since at $B=0$ no spin-electric coupling can be obtained at
the first order of the spin-orbit interaction (see further).

In Eq.~(\ref{HamH0}), $H_{SO}$ stands for the spin-orbit Hamiltonian.
We start with considering the so-called ``linear in $p$'' spin-orbit interaction,
\begin{equation}
H_{SO}=\alpha(p_x\sigma_y-p_y\sigma_x)+\beta(-p_x\sigma_x+p_y\sigma_y),
\label{HSO}
\end{equation}
which is the sum of the Rashba ($\alpha$)~\cite{Rashba} and 
2D Dresselhaus ($\beta$)~\cite{DyakKach,Dress} spin-orbit interactions.
This type of spin-orbit interaction gives rise to a sizable phonon-induced spin relaxation rate $1/T_1$,~\cite{GKL,KhaNaz} 
of the same order of magnitude as experimentally measured.~\cite{ENature,Kroutvar,AmashaZumbuhl}
Moreover, in the 2D limit the linear in $p$ spin-orbit interaction is dominant, because $\beta\propto 1/d^2$.

In the standard EDR scheme, an alternating magnetic field is generated by a
current in a nearby conductor. 
In our setup (see Fig.~\ref{EDSRfigure}), no charge flow is ideally present between the gates.
However, the alternating electric field $\mbox{\boldmath $E$}({\bm r},z,t)$ gives rise to a 
displacement current, with the current density
\begin{equation}
{\bm J}_D({\bm r},z,t)=\frac{\kappa}{4\pi}\frac{\partial\mbox{\boldmath $E$}({\bm r},z,t)}{\partial t},
\end{equation}
where $\kappa$ is the electric permittivity.
The external magnetic field $\mbox{\boldmath $B$}$ acquires, thus, an {\em ac}-component, 
$\mbox{\boldmath $B$}\to \mbox{\boldmath $B$}+\mbox{\boldmath ${\cal B}$}(t)$,
where
$\mbox{\boldmath ${\cal B}$}(t)=\mbox{\boldmath $\nabla$}\times \mbox{\boldmath ${\cal A}(t)$}$.
The vector potential $\mbox{\boldmath ${\cal A}$}({\bm r},z,t)$ is obtained as usual from
Ampere's law~\cite{Jackson}
\begin{equation}
\nabla^2\mbox{\boldmath ${\cal A}$}=-\frac{4\pi\mu}{c}\left({\bm J}+{\bm J}_D\right)
,\label{eqAmperslaw}
\end{equation}
where $\mu$ is the magnetic permittivity and ${\bm J}$ is the charge flow density (in our case $J=0$).
In Eq.~(\ref{eqAmperslaw}), we adopted the Coulomb gauge $\mbox{\boldmath $\nabla$}\cdot\mbox{\boldmath ${\cal A}$}=0$ and 
used the notation $\mbox{\boldmath $\nabla$}=(\partial/\partial{\bm r},\partial/\partial z)$.

The magnetic field $\mbox{\boldmath ${\cal B}$}(t)$ couples to the electron spin via the
Zeeman interaction in Eq.~(\ref{HZ}) [with ${\bm B}\to {\bm B}+\mbox{\boldmath ${\cal B}$}(t)$],
giving rise to an ESR source, which can be used, in principle, for spin manipulation 
in quantum dots.
However, the amplitude of $\mbox{\boldmath ${\cal B}$}(t)$ is, in practice,
extremely small; it is proportional to $1/c$, 
as expected from the relativistic nature of $\mbox{\boldmath ${\cal B}$}(t)$.
Furthermore, the proximity of the top gates to the 2DEG decrease the displacement current
enclosed by the magnetic field lines penetrating the quantum dot.
Using Eq.~(\ref{eqAmperslaw}), we estimate ${\cal B}=\mu\kappa c^{-1} L_z\partial E/\partial t\approx 10^{-6}\,{\rm G}$,
for a quantum dot which is $L_z=100\,{\rm nm}$ below the gates plane and an electric field
$E(t)=E_0\sin(\omega_{\it ac}t)$, with amplitude $\mu\kappa E_0=10^2\,{\rm V}/{\rm cm}$
and frequency $\omega_{ac}/2\pi=10^9\,{\rm Hz}$.
As we show below, a much stronger effective magnetic field arises from the EDSR effect in
the present setup, and therefore, the displacement current can be safely ignored.

Recently, a sizable ESR effect has been obtained with the help of a wire
placed on top of a GaAs double dot.~\cite{noteDelft}
In this case, $J\neq 0$ and the magnitude of $\mbox{\boldmath ${\cal B}$}(t)$ is estimated
from Eq.~(\ref{eqAmperslaw}) to be 
${\cal B}=\pi\mu c^{-1}I/(L_y+L_z)$, 
where $I$ is current in the wire and $2L_y$ is the lateral size of the wire.
For $L_y=L_z=100\,{\rm nm}$ and $I=1\,{\rm mA}$, the magnetic field obtained in this setup 
is on the order of ${\cal B}\sim 10\,{\rm G}$.

\section{Spin-electric coupling}
\label{secSpElCo}
Now, we focus on the electric-field component of the {\em ac}-signal
and show that, together with the spin-orbit interaction $H_{SO}$ and Zeeman splitting $H_Z$,
it suffices to generate a sizable EDSR field in the quantum dot.
For simplicity, we set $\mbox{\boldmath ${\cal B}$}(t)\to 0$ from now on and choose 
$\mbox{\boldmath ${\cal A}$}(t)=0$. 
As a result, we retain only a constant in space and time magnetic field 
${\bf B}=B(\cos\varphi\sin\vartheta,\sin\varphi\sin\vartheta,\cos\vartheta)$ 
and an alternating electric field ${\bm E}({\bm r},t)=(1/e)\mbox{\boldmath $\nabla$}V({\bm r},t)$.
Thus, we consider further the Hamiltonian in Eq.~(\ref{HamH}), assuming that $H_0$ is time-independent
and describes the dot in the absence of {\em ac} fields.

We aim at diagonalizing $H_0$ using a Schrieffer-Wolff transformation, similar to Ref.~\onlinecite{GKL}.
We thus look for a transformation matrix $S$ 
such that the transformed Hamiltonian $\tilde H_0=\exp(S)H_0\exp(-S)$ is fully diagonal, see Appendix~\ref{appSW}.
At $B=0$, the ground state of $H_0$ (and also of $\tilde H_0$) is a Kramers doublet, 
because the spin-orbit interaction is always time-reversal symmetric (at $B=0$).
We therefore choose to encode the qubit into  
the ground state Kramers doublet of the quantum dot.
Owing to the mixed spin and orbital nature of the states
an alternating potential $V({\bm r},t)$, such as in Eq.~(\ref{HamH}), couples to the qubit.
We proceed to derive this coupling, by calculating the transformation matrix $S$
at the leading order of spin-orbit interaction, 
\begin{equation}\label{SmatrixHSO}
S=\frac{1-{\cal P}}{\hat{L}_d+\hat{L}_Z}H_{SO}+{\cal O}(H_{SO}^2),
\end{equation}
where $\hat{L}_d$ and $\hat{L}_Z$ are Liouville superoperators, {\em i.e.} 
$\hat{L}_dA=[H_d,A]$ and $\hat{L}_ZA=[H_Z,A]$, $\forall A$.
The projector ${\cal P}$ projects onto the 
diagonal (or degenerate) part of the Hilbert space of $H_d+H_Z$, which ensures applicability
of ``non-degenerate'' perturbation theory. 
The coupling of spin to electric fields is then found by applying the same Schrieffer-Wolff transformation 
to the potential $V({\bm r},t)$.
We obtain the following effective Hamiltonian for our qubit in the presence of an alternating potential $V({\bm r},t)$,
to leading order in the spin-orbit interaction,
\begin{equation}\label{HeffHZSV}
H_{\rm eff}=H_Z+\langle\psi_0|[S,V({\bm r},t)]|\psi_0\rangle,
\end{equation}
where $S$ is the transformation matrix in Eq.~(\ref{SmatrixHSO}) and
$|\psi_0\rangle$ is the quantum dot ground state.
For a quantum dot with a harmonic confining potential, $U({\bm r})=m_e\omega_0^2r^2/2$,
the transformation matrix $S$ was calculated in Ref.~\onlinecite{BGL} to all orders 
of the Zeeman interaction and the first order of the ``linear in $p$'' spin-orbit interaction~(\ref{HSO}). 
For simplicity, we consider here only the linear in $B$ terms,
\begin{eqnarray}\label{Smatrixxizeta}
S&=&i\mbox{\boldmath $\xi$}\cdot\mbox{\boldmath $\sigma$}-
\frac{E_Z}{m_e\omega_0^2}[{\bm n}\times\mbox{\boldmath $\zeta$}]\cdot\mbox{\boldmath $\sigma$},\\
\mbox{\boldmath $\xi$}&=&(\lambda_-^{-1}y',\lambda_+^{-1}x',0),\nonumber\\
\mbox{\boldmath $\zeta$}&=&(\lambda_-^{-1}\partial/\partial y',\lambda_+^{-1}\partial/\partial x',0),\nonumber
\end{eqnarray}
where $\lambda_\pm=\hbar/m_e(\beta\pm\alpha)$ are the spin-orbit lengths, and the
vectors $\mbox{\boldmath $\xi$}$ and $\mbox{\boldmath $\zeta$}$ are given in the
coordinate frame $x'=(x+y)/\sqrt{2}$, $y'=-(x-y)/\sqrt{2}$, and $z'=z$ (see Fig.~\ref{EDSRfigure}).
The first term in Eq.~(\ref{Smatrixxizeta}) commutes with scalar potentials
and therefore drops out in Eq.~(\ref{HeffHZSV}).
More generally, for arbitrary confining potential, the first term 
is replaced by $i(1-{\cal P})\mbox{\boldmath $\xi$}\cdot\mbox{\boldmath $\sigma$}$,
resulting nevertheless in no coupling of spin to electric fields. 
The second term in Eq.~(\ref{Smatrixxizeta}), however, allows us to express the coupling of spin to charge via the
electric field $\mbox{\boldmath $E$}(t)=(1/e)\langle\psi_0|\mbox{\boldmath $\nabla$}V({\bm r},z,t)|\psi_0\rangle$
that acts on the quantum dot electron.
For the harmonic confining potential, we obtain 
\begin{eqnarray}\label{Heffexpl}
H_{\rm eff}&=&\frac{1}{2}g\mu_B{{\bm B}}\cdot\mbox{\boldmath $\sigma$}+\frac{1}{2}{\bm h}(t)\cdot\mbox{\boldmath $\sigma$},\\
{\bm h}(t)&=&2g\mu_B{\bm B}\times\mbox{\boldmath $\Omega$}(t),\label{hoftdef}\\
\mbox{\boldmath $\Omega$}(t)&=&\frac{-e}{m_e\omega_0^2}\left(\lambda_-^{-1}E_{y'}(t),\lambda_+^{-1}E_{x'}(t),0\right).
\label{Omegaexpl}
\end{eqnarray}
The dimensionless field $\mbox{\boldmath $\Omega$}(t)$ describes a combined effect of the spin-orbit interaction
and electric fields (or more generally potential fluctuations) on the qubit.
$\mbox{\boldmath $\Omega$}(t)$ was calculated in Ref.~\onlinecite{GKL} for the phonon potential and
in Ref.~\onlinecite{BGL} for the shot-noise of a QPC nearby the quantum dot.
In our case, $\mbox{\boldmath $\Omega$}(t)$ is merely a classical driving field generated by the
{\em ac}-signal.

Considering further a constant in space (at least on the scale of the quantum dot), 
alternating electric field 
$\mbox{\boldmath $E$}(t)=\mbox{\boldmath $E$}_0\sin(\omega_{\it ac}t)$
of amplitude $\mbox{\boldmath $E$}_0=E_0(\cos\phi,\sin\phi,0)$,
where $\phi$ is the angle of $\mbox{\boldmath $E$}_0$ with respect to the axis $x'$ (see Fig.~\ref{EDSRfigure}),
we obtain explicitly $\mbox{\boldmath $\Omega$}(t)=\mbox{\boldmath $\Omega$}_0\sin(\omega_{\it ac}t)$, with
\begin{equation}
\mbox{\boldmath $\Omega$}_0=\frac{-eE_0}{m_e\omega_0^2}
\left(\lambda_-^{-1}\sin\phi,\lambda_+^{-1}\cos\phi,0\right).
\end{equation}
To give an estimate for the amplitude $\Omega_0$ in GaAs quantum dots, we assume 
$\lambda_+\approx\lambda_-\approx\lambda_{SO}=8\,\mu{\rm m}$, $\hbar\omega_0=1\,{\rm meV}$, 
and $E_0=10^2\,{\rm V}/{\rm cm}$, which yields $\Omega_0\sim 10^{-3}$.

The amplitude of the resulting effective magnetic field due to EDSR is found from Eq.~(\ref{hoftdef}) 
to be 
\begin{equation}
\delta{\bm B}_0=2{\bm B}\times\mbox{\boldmath $\Omega$}_0.
\label{esdB02BO}
\end{equation}
The maximal amplitude is obtained for $\mbox{\boldmath $B$}\perp\mbox{\boldmath $\Omega$}_0$,
which in experiment can easily be arranged for by, e.g., choosing $\mbox{\boldmath $B$}\parallel z$.
In-plane magnetic fields can also be used, provided $\mbox{\boldmath $E$}(t)$ is linearly polarized.
For example, an electric field $\mbox{\boldmath $E$}(t)$ aligned with $x'$ generates, 
according to Eq.~(\ref{Omegaexpl}), a dimensionless field $\mbox{\boldmath $\Omega$}(t)$ along $y'$.
In this case, $\mbox{\boldmath $B$}$ should be chosen along $x'$ for maximal spin-electric coupling.

Using our previous estimate for $\Omega_0\sim 10^{-3}$, we obtain from Eq.~(\ref{esdB02BO}) that
$\delta B_0\sim 1\,{\rm mT}$ for a magnetic field $B=1\,{\rm T}$ oriented transversely 
to $\mbox{\boldmath $\Omega$}_0$.
In principle, the dimensionless factor $\Omega_0$ can be increased up to $\Omega_0\sim 1$.
However, this requires a specially designed setup, where the value of the
electron displacement $r_0$ in Eq.~(\ref{eqr0}) approaches the spin-orbit length $\lambda_{SO}$.

Next we remark that $\mbox{\boldmath $\Omega$}(t)$ in Eq.~(\ref{Omegaexpl}) 
can be written by the order of magnitude as $\Omega(t)\sim r_0(t)/\lambda_{SO}$.
More rigorously, we rewrite Eq.~(\ref{Omegaexpl}) in the following form
\begin{equation}
\Omega_i(t)=\sum_j\left(\lambda_{SO}^{-1}\right)_{ij}r_{0j}(t),
\label{eqOittensorform}
\end{equation}
where $\left(\lambda_{SO}^{-1}\right)_{ij}$ is a tensor of inverse spin-orbit lengths,
\begin{equation}
\left(\lambda_{SO}^{-1}\right)_{ij}=\left(
\begin{array}{ccc}
0&1/\lambda_-\\
1/\lambda_+&0
\end{array}
\right),\label{eq1lamSOtensorpm}
\end{equation}
with $1/\lambda_\pm=m_e(\beta\pm\alpha)/\hbar$ and the frame $(x',y')$ was used to
represent the tensor. 
For order of magnitude estimates, it is useful to introduce
the scalar
\begin{equation}
\frac{1}{\lambda_{SO}}=\frac{1}{\sqrt{2}}\| \lambda_{SO}^{-1}\|,
\end{equation}
where $\|\lambda_{SO}^{-1}\|$ is the Frobenius norm of $(\lambda_{SO}^{-1})_{ij}$.
In the case of Eq.~(\ref{eq1lamSOtensorpm}), we have
$1/\lambda_{SO}=(m_e/\hbar)\sqrt{\alpha^2+\beta^2}$.

Despite the fact that Eq.~(\ref{eqOittensorform}) was obtained considering
the harmonic confining potential as an example, its generality suggests
that it should remain valid for quantum dots of arbitrary confinement,
provided, to a good approximation, the {\em ac}-signal merely displaces the
quantum dot parallel to itself by a vector ${\bm r}_0(t)$ as a function of time.
Note that ${\bm r}_0(t)$ is the only available parameter to be compared with $\lambda_{SO}$
in the limit of strong confinement ($\lambda_d\to 0$).
We extend the class of Hamiltonians considered here to
any combination of confinement and {\em ac}-voltage potential that can be
rewritten in the form
\begin{equation}
U({\bm r})+V({\bm r},t)=U\left({\bm r}-{\bm r}_0(t)\right)+V_0(t),
\label{UVeqUV}
\end{equation} 
where $V_0(t)$ is independent of ${\bm r}$.
We note that, as before, the electron wave function extension $\lambda_d$
is assumed to be small compared to the spin-orbit length
$\lambda_{SO}$ at each moment in time.
Equation (\ref{UVeqUV}) need not be satisfied exactly.
Note that $\lambda_d$ enters only in the definition of ${\bm r}_0(t)$ and does
appear alone as a parameter in Eq.~(\ref{eqOittensorform}).
Therefore, defining ${\bm r}_0(t)$ as the average electron position, 
${\bm r}_0(t)=\int {\bm r}\left|\psi({\bm r},t)\right|^2d^2r$,
we expect Eq.~(\ref{eqOittensorform}) to be valid to leading order also when
the electron probability density $\left|\psi({\bm r},t)\right|^2$ changes shape, 
but the dot size changes weakly.

Equations~(\ref{Heffexpl}), (\ref{hoftdef}), and (\ref{eqOittensorform}) 
form the basis of EDSR in quantum dots and can be used to efficiently manipulate the 
electron spin by electrical gates.
Finally, we remark that Eqs.~(\ref{Heffexpl})-(\ref{Omegaexpl}) have been derived
under the assumption $r_0\ll\lambda_{SO}$, and therefore, can be used only for 
$\Omega_{0}\ll 1$. 
In Sec.~\ref{secDisc} we discuss the case of $\Omega_{0}\sim 1$ in more detail.
A further assumption in deriving Eqs.~(\ref{Heffexpl})-(\ref{Omegaexpl}) was that
the frequency spectrum of ${\bm E}(t)$ lies well below the size-quantization energy 
$\hbar\omega_0$. 
This adiabaticity constraint is generic to the spin-based quantum 
computation~\cite{ALS,LD}; it guarantees that the electron is not excited 
to higher in energy orbital levels.

\section{Spin dynamics and coherence}
\label{secDynDec}
The electron spin obeys the Bloch equation~\cite{Bloch,CohenTannoudji}
\begin{equation}
\langle\dot{\bm S}\rangle=[{\bm\omega}_Z+\delta{\bm\omega}(t)]\times\langle{\bm S}\rangle
-\Gamma\langle{\bm S}\rangle + {\bm \Upsilon},
\label{bloch}
\end{equation}
where ${\bm\omega}_Z=g\mu_B{\bm B}/\hbar$ is the Larmor spin-precession frequency and
$\delta{\bm\omega}(t)={\bm h}(t)/\hbar$.
The spin relaxation tensor $\Gamma_{ij}$ and the inhomogeneous part $\Upsilon_i$
are due to the environment and can be derived microscopically~\cite{Bloch,GKL} 
within the Born-Markov approximation.
Strictly speaking, $\Gamma_{ij}$ and $\Upsilon_i$ in Eq.~(\ref{bloch})
depend also on the driving.
In particular, $\Gamma_{ij}$ acquires, in general, a time-dependent part.
However, we neglect these effects here since the energy scales are well separated.
Indeed, from experiments~\cite{ENature,Kroutvar,AmashaZumbuhl} and theory,~\cite{GKL,KhaNaz} we infer that
$\Gamma_{ij},\Upsilon_i\sim (10^2-10^6)\,{\rm s}^{-1}$, {\em i.e.} they are very small, 
so that the regime $\Gamma_{ij},\Upsilon_i\ll \delta \omega\ll \omega_Z$ holds.
In this regime, the rotating wave approximation~\cite{CohenTannoudji} is valid.
We consider a completely general driving field
\begin{equation}
\delta{\bm\omega}(t)=\delta{\bm\omega}_a\sin(\omega_{\it ac}t)+\delta{\bm\omega}_b\cos(\omega_{\it ac}t),
\end{equation}
which can be realized in practice by implementing two independent electric fields at the quantum dot site.
This is, however, by no means necessary for our EDSR scheme.

The Rabi frequency then reads
\begin{equation}
{\bm\omega}_R=\frac{1}{2}\left(\delta{\bm\omega}_a\times{\bm n}
-\left[\delta{\bm\omega}_b\times{\bm n}\right]\times{\bm n}\right).
\label{eqnomegaRabi}
\end{equation}
Here, we assume that $\omega_{\it ac}$ is not far from resonance, 
{\em i.e.} $\left|\omega_{\it ac}-\omega_Z\right|<\omega_{\it ac}/2$.
In a coordinate frame $(X,Y,Z)$ with $Z\parallel {\bm B}$,
the spin dynamics is approximated as follows
\begin{eqnarray}
\langle S_\pm(t)\rangle&\approx&\tilde{S}_\pm(t)e^{\pm i\omega_{\it ac}t}\\
\langle S_Z(t)\rangle&\approx&\tilde{S}_Z(t),
\end{eqnarray} 
where $S_\pm=S_X\pm iS_Y$. 
The spin $\tilde{\bm S}(t)$ obeys a simpler (static) Bloch equation 
\begin{equation}
\dot{\tilde{\bm S}}=\left({\bm\delta}+{\bm\omega}_R\right)\times\tilde{\bm S}
-\tilde{\Gamma}\tilde{\bm S} + \tilde{\bm \Upsilon},
\label{tildeSBloch}
\end{equation}
where
${\bm \delta}=({\omega}_Z-\omega_{\it ac}){\bm n}$ gives the detuning from resonance.
The relaxation tensor $\tilde\Gamma_{ij}$ is diagonal, with
$\tilde\Gamma_{XX}=\tilde\Gamma_{YY}=1/T_2$ and $\tilde\Gamma_{ZZ}=1/T_1$, and
$\tilde{\Upsilon}_i$ assumes $\tilde{\Upsilon}_i=\tilde{\Gamma}_{ij} S_j^T$.
Here, $T_1$ and $T_2$ are the relaxation and decoherence times in the absence of driving 
measured in experiment,~\cite{ENature,Kroutvar,AmashaZumbuhl} and ${\bm S}^T=-({\bm n}g/2|g|)\tanh(E_Z/2k_BT)$
is the thermodynamic value of spin, with $T$ being the temperature.

The time-evolution of $\tilde{\bm S}$ in Eq.~(\ref{tildeSBloch}) is simplest in
a coordinate frame $(X',Y',Z')$, with $Z'\parallel (\mbox{\boldmath $\delta$}+\mbox{\boldmath $\omega$}_R)$,
and reads
\begin{eqnarray}
\tilde{S}_{X'}(t)&=&S^0_\perp e^{-t/\tilde T_2}\sin\left(t\sqrt{\delta^2+\omega_R^2}+\phi\right),\nonumber\\
\tilde{S}_{Y'}(t)&=&S^0_\perp e^{-t/\tilde T_2}\cos\left(t\sqrt{\delta^2+\omega_R^2}+\phi\right),\nonumber\\
\tilde{S}_{Z'}(t)&=&\tilde{S}_T+(S^0_{Z'}-\tilde{S}_T)e^{-t/\tilde T_1},
\end{eqnarray}
where $S^0_\perp$, $S^0_{Z'}$, and $\phi$ give the initial spin state, 
$\langle{\bm S}(0)\rangle\equiv\tilde{\bm S}(0)=(S^0_\perp\sin\phi,S^0_\perp\cos\phi,S^0_{Z'})$, 
in the coordinate frame $(X',Y',Z')$.
Furthermore, the decay times $\tilde{T}_1$ and $\tilde{T}_2$ read
\begin{eqnarray}
\frac{1}{\tilde{T}_1}&=&\frac{1}{\delta^2+\omega_R^2}
\left(\frac{\delta^2}{T_1}+\frac{\omega_R^2}{T_2}\right),\nonumber\\
\frac{1}{\tilde{T}_2}&=&\frac{1}{2(\delta^2+\omega_R^2)}
\left(\frac{\omega_R^2}{T_1}+\frac{2\delta^2+\omega_R^2}{T_2}\right).
\end{eqnarray}

The stationary value of spin 
$\tilde{\bm S}_T:=\tilde{\bm S}(t\to\infty)$ to leading order reads
\begin{equation}
\tilde{\bm S}_T=-\frac{g}{2|g|}
\frac{(\mbox{\boldmath $\delta$}+\mbox{\boldmath $\omega$}_R)\delta}{\delta^2+(T_1/T_2)\omega_R^2}
\tanh\left(E_Z/2k_BT\right).
\label{STtildeleading}
\end{equation}
Note that at resonance ($\delta=0$), the right-hand side in Eq.~(\ref{STtildeleading}) vanishes.
Therefore, in the vicinity of resonance, $\tilde{\bm S}_T$ is determined by the subleading order term, 
which can be obtained from Eq.~(\ref{STtildeleading}) by 
replacing the numerator 
$(\mbox{\boldmath $\delta$}+\mbox{\boldmath $\omega$}_R)\delta\to(1/T_2)[{\bm\omega}_R\times{\bm n}]$.
Measurement of $\tilde{\bm S}_T$ in the presence of driving provides information about the spin lifetimes $T_{1,2}$.
For instance, at resonance the relaxation time $T_1$ can be accessed at the leading order of 
$\Gamma_{ij}/\omega_R\ll 1$,
\begin{equation}
\tilde{\bm S}_T\left(\delta=0\right)=-\frac{g}{|g|}
\frac{{\bm\omega}_R\times{\bm n}}{2T_1\omega_R^2}
\tanh\left(E_Z/2k_BT\right).
\label{STtildeleadingb}
\end{equation}

Finally, we estimate the Rabi frequency $\omega_R$ using Eq.~(\ref{eqnomegaRabi}) and the parameters from 
Sec.~\ref{secSpElCo}. 
For $\Omega_0\sim 10^{-3}$, $|g|=0.44$, and $B=10\,{\rm T}$ we obtain
$\omega_R\sim 10^{8}\,{\rm s}^{-1}$.
We conclude that, with the present quantum-dot setups, EDSR enables one to manipulate the electron spin 
on a time scale of $10\,{\rm ns}$, which is considerably shorter than
the spin lifetimes, for which values between $1$ and  $150\;{\rm ms}$ 
(depending on the applied magnetic field) in gated GaAs quantum dots have been reported
recently.~\cite{ENature,AmashaZumbuhl}

\section{$p^3$-Dresselhaus terms}
\label{secSOp3}

Next we consider the so-called $p^3$ terms 
of the Dresselhaus spin-orbit interaction~\cite{DyakKach},
\begin{equation}
H_{SO}=\frac{\gamma}{2}\left(p_yp_xp_y\sigma_x-p_xp_yp_x\sigma_y\right),
\label{HSOp3}
\end{equation}
where $\gamma=\alpha_c/\sqrt{2m_e^3E_g}$ is the spin-orbit coupling constant,
with $\alpha_c$ ($\approx 0.07$ for GaAs~\cite{DyakKach}) being a dimensionless constant 
defined in Ref.~\onlinecite{PikusTitkov} and $E_g$ the band gap.
For simplicity, we impose here the dipolar approximation for the
{\em ac}-signal,
\begin{equation}\label{Vdipolarapprox}
V(\mbox{\boldmath $r$},t)=e\int_{0}^{{\bm r}} d\mbox{\boldmath $r$}'\cdot\mbox{\boldmath $E$}(\mbox{\boldmath $r$}',t)
\approx e\mbox{\boldmath $E$}(t)\cdot\mbox{\boldmath $r$}.
\end{equation}
Quite remarkably, 
if the quantum dot potential is harmonic,
$U({\bm r})=\sum_{ij}u_{ij}r_ir_j$,
then the spin does not couple to $\mbox{\boldmath $E$}(t)$ 
at the first order of $H_{SO}$ and zeroth order of $E_Z$.
Indeed, the second term in Eq.~(\ref{HeffHZSV}) vanishes
for $V({\bm r},t)=e\mbox{\boldmath $E$}(t)\cdot\mbox{\boldmath $r$}$ 
and $S=\hat{L}_d^{-1}H_{SO}$
because of the following two identities
\begin{eqnarray}
\langle\psi_0|[\hat{L}_d^{-1}H_{SO},{\bm r}]|\psi_0\rangle&=&\langle\psi_0|[\hat{L}_d^{-1}{\bm r},H_{SO}]|\psi_0\rangle,\\
{[\hat{L}_d^{-1}{\bm r},H_{SO}]}&=&0,\;\;\;\;\;\forall\; H_{SO}({\bm p}).
\end{eqnarray}
The latter is specific to $H_d$ in Eq.~(\ref{Hdot}) with a harmonic $U({\bm r})$, for which
the operator $\hat{L}_d^{-1}{\bm r}$ can be expressed via the components of ${\bm p}-(e/c){\bm B}_z\times{\bm r}$.
Note that, generally, $[{\bm p},H_{SO}]=(e/c)[{\bm B}_z\times{\bm r},H_{SO}]$ for any $H_{SO}$ that is a function
of only ${\bm p}=(p_x,p_y)$. 
Thus, for a harmonic confining potential, one is left with the same dominant mechanism as
considered above for the "linear in $p$" terms.
Expanding in terms of the Zeeman interaction, we recover Eqs.~(\ref{Heffexpl}) and (\ref{hoftdef}) 
with $\mbox{\boldmath $\Omega$}(t)$ given now by
\begin{eqnarray}
&&\Omega_i(t)=-\frac{e}{m_e\omega_0^2}\sum_j{\left(\lambda_{SO}^{-1}\right)}_{ij}E_j(t),\\
&&{\left(\lambda_{SO}^{-1}\right)}_{ij}=\frac{m_e}{\hbar}\langle\psi_0|\frac{\partial^2H_{SO}}{\partial\sigma_i\partial p_j}|\psi_0\rangle,
\end{eqnarray}
where $(\lambda_{SO}^{-1})_{ij}$ is a tensor of inverse spin-orbit lengths, 
and as before we consider $U({\bm r})=m_e\omega_0^2r^2/2$.
For $H_{SO}$ in Eq.~(\ref{HSOp3}), we obtain explicitly $(\lambda_{SO}^{-1})_{ij}=\frac{1}{4}\gamma\omega_0m_e^2\delta_{ij}$
and $\Omega_i(t)=-\gamma em_eE_i(t)/4\omega_0$.
To estimate the strength of the resulting EDSR, we note that $\gamma\sim\beta d^2/\hbar^2$, and 
therefore the amplitude of $\mbox{\boldmath $h$}(t)=2g\mu_B{\bm B}\times\mbox{\boldmath $\Omega$}(t)$
is down now by a factor $d^2/\lambda^2\ll 1$ compared to the $p$-terms.

Next we consider a quantum dot with non-harmonic potential $U({\bm r})$ and show that
the $p^3$-terms in Eq.~(\ref{HSOp3}) give rise to a spin-electric coupling
proportional to the cyclotron frequency $\omega_c=eB_z/m_ec$.
Since $\hbar\omega_c$ differs parametrically from $E_Z$
($E_Z/\hbar\omega_c=gm_eB/2mB_z$), the $p^3$-terms can be as
significant as the $p$-terms, provided $E_Z/\hbar\omega_c\lesssim d^2/\lambda^2$, which is
realistic for GaAs quantum dots.
Note that for the $p$-terms no spin-electric coupling proportional to $\omega_c$ arises
at the first order of $H_{SO}$.
We thus leave out $\hat{L}_Z$ in Eq.~(\ref{SmatrixHSO}) and consider
a confining potential $U({\bm r})$ that differs from
a harmonic one by a function $W({\bm r})$,
\begin{equation}
U({\bm r})=\sum_{ij}u_{ij}r_ir_j+W({\bm r})\equiv U_H({\bm r})+W({\bm r}),
\end{equation}
where $u_{ij}$ are real coefficients and $W({\bm r})={\cal O}(r^3)$.
While in general $W({\bm r})$ need not be small compared to $H_H=p^2/2m_e+U_H({\bm r})$, 
in the following we expand the denominator of Eq.~(\ref{SmatrixHSO}) in terms of $W\ll H_H$, 
considering therefore only small deviations of $U({\bm r})$ from harmonic potentials.
Then, using Eqs.~(\ref{HeffHZSV}), (\ref{Heffexpl}), and (\ref{Vdipolarapprox}), we obtain
at leading order in $\omega_c$ 
\begin{equation}\label{hiofowc}
\frac{h_i(t)}{\omega_c}=e{\bm E}(t)\cdot
\left.\langle\psi_0|[{\bm R}({\bm r}),\frac{\partial^2 S_H}
{\partial\omega_c\partial\sigma_i}]|\psi_0\rangle\right|_{\omega_c=0}, 
\end{equation}
where we set $\omega_c\to0$ in the right-hand side of Eq.~(\ref{hiofowc}) 
after evaluating $\partial S_H/\partial\omega_c$ in the symmetric gauge, with
$S_H$ defined as $[H_H,S_H]=H_{SO}$. 
The linear relationship between $h_i$ and $\omega_c$ holds for $\omega_c\ll\omega_0$,
where $\omega_0\equiv 2\sqrt{\det(u)/m_e{\rm Tr}(u)}$.
In Eq.~(\ref{hiofowc}), the perturbation $W({\bm r})$ enters via the function ${\bm R}({\bm r})$
defined as follows
\begin{equation}
{R}_i({\bm r})=\sum_j{\left(u^{-1}\right)}_{ij}\frac{\partial W({\bm r})}{\partial r_j}.
\end{equation} 
Note that $\langle R/r\rangle\sim W_0\lambda_d/\hbar\omega_0\lambda_W$
is the small parameter of our expansion in terms of $W({\bm r})$, with 
$W_0$ and $\lambda_W\lesssim\lambda_d$ being, respectively, the characteristic amplitude and length scale
of the variation of $W({\bm r})$ over the quantum dot size.
It is important to note that the antisymmetric part of $W({\bm r})$ drops out in
Eq.~(\ref{hiofowc}) because $H_{SO}$ is also antisymmetric with respect to ${\bm r}\to -{\bm r}$.

Next, as an example, we consider $U_H({\bm r})=m_e\omega_0^2r^2/2$ and $W({\bm r})=\eta r^4$,
and obtain
\begin{equation}
\frac{1}{2}{\bm h}(t)\cdot\mbox{\boldmath $\sigma$}=\frac{e\gamma\eta\hbar^2\omega_c}{9m_e\omega_0^4}
\left(E_y(t)\sigma_x+E_x(t)\sigma_y\right).
\label{eq12htseqEE}
\end{equation}
Here, we have used the deformation quantization theory,~\cite{DefQuant} which
allowed us to considerably simplify the derivation of Eq.~(\ref{eq12htseqEE}) by
performing most of the calculation in classical mechanics and only at the
final stage come back to quantum mechanics. 
We have also carried out a fully quantum derivation of Eq.~(\ref{eq12htseqEE}) 
and recovered the same result.

To estimate the strength of the resulting EDSR, we note that
$h\sim\hbar\omega_c(\lambda_d/\lambda_{SO})(e\lambda_d E_0/\hbar\omega_0)\langle R/r\rangle$, where 
$\lambda_{SO}=4/\gamma\omega_0m_e^2$ ($\approx\lambda_d^2/[0.01\,{\rm nm}]$ for GaAs) 
is the spin-orbit length of the $p^3$-terms
and the parameter $\langle R/r\rangle\sim W(\lambda_d)/\hbar\omega_0$ characterizes the deviation 
of the quantum dot confinement from harmonic. 
In practice, $\langle R/r\rangle$ can be as large as unity, but here we assume 
$\langle R/r\rangle\sim\eta\lambda_d^4/\hbar\omega_0=0.1$.
For an electric field with amplitude $E_0=10^2\,{\rm V}/{\rm cm}$ and 
a GaAs quantum dot with $\hbar\omega_0=1\,{\rm meV}$, 
we obtain the equivalent of an {\em ac} magnetic field $\delta {\bm B}(t)={\bm h}(t)/g\mu_B$
that has an amplitude $\delta B_0\approx 1\,{\rm mT}$ at $B_z=1\,{\rm T}$ and $|g|=0.44$.
In contrast to the previous mechanism, $\delta {\bm B}(t)$ can have here also a finite longitudinal component
$\delta{\bm B}_\parallel(t)={\bm n}({\bm n}\cdot\delta {\bm B}(t))$, which however vanishes if ${\bm B}\parallel z$.

Finally, we note that the $p^3$-terms can also be relevant for spin relaxation
in quantum dots with non-harmonic confining potential.
Of course, the magnetic field has to have an out-of-plane
component for this spin-electric coupling to dominate
over the one considered in Sec.~\ref{secSpElCo}.

\section{Discussions}
\label{secDisc}
The coupling of spin to electric fields that we have derived above can be used in a variety 
of ways to access and manipulate the electron spin in experiments.
The effective Hamiltonian in Eq.~(\ref{Heffexpl}) 
has the same form as the Hamiltonian of an ESR effect.
This shows that ESR and EDSR are mutually interchangeable and 
the choice of the effect to be used depends on the particular experimental setup.
In GaAs quantum dots, the spin-orbit interaction is weak enough to ensure long coherence times
and, at the same time, strong enough to allow room for spin manipulation on 
an experimentally accessible timescale of $\sim 10\;{\rm ns}$.
Much shorter timescales can be achieved in InAs quantum dots, because of a much
stronger spin-orbit coupling and a larger electron $g$-factor. 
In contrast, the EDSR effect is of little use in materials with very weak or nearly absent
spin-orbit interaction, such as, e.g., carbon-nanotube quantum dots.
To make order-of-magnitude estimates easier,
we draw an analogy between the EDSR and ESR effects in terms of 
the particular way the ${\bm B}$ and ${\bm E}$ fields couple to the electron spin
${\bm S}=(\hbar/2)\mbox{\boldmath $\sigma$}$.

We recall that the ESR effect occurs as a result of the Zeeman interaction
of the electron spin with an {\em ac} magnetic field.
It is convenient to write this interaction in the form of a
magnetic dipole interaction,
\begin{equation}
H_{\rm ESR}=-\mbox{\boldmath $\mu$}\cdot\mbox{\boldmath $B$}(t),
\label{HESRmuBoft}
\end{equation} 
where $\mbox{\boldmath $B$}(t)$ is the {\em ac} magnetic field and
$\mbox{\boldmath $\mu$}$ is the electron magnetic moment,
\begin{equation}
\mbox{\boldmath $\mu$}=-\frac{1}{2}g\mu_B\mbox{\boldmath $\sigma$},
\label{defofmugmuBsigma}
\end{equation}
where $g$ is, in general, a tensor, see Eq.~(\ref{renormZeemangijsB}).

By analogy with the ESR effect, the EDSR effect can be viewed
as arising from an interaction between the {\em ac} electric field 
$\mbox{\boldmath $E$}(t)$ and a spin-electric moment $\mbox{\boldmath $\nu$}$.
The respective spin-electric interaction is then analogous to Eq.~(\ref{HESRmuBoft})
and reads
\begin{equation}
H_{\rm EDSR}=-\mbox{\boldmath $\nu$}\cdot\mbox{\boldmath $E$}(t),
\label{HEDSRnuEoft}
\end{equation}
where the spin-electric moment $\mbox{\boldmath $\nu$}$ is due to
an interplay between the spin-orbit interaction 
and some time-reversal breaking interaction, such as the Zeeman interaction.
This analogy is not complete.
Equation~(\ref{HEDSRnuEoft}) is valid only for {\em ac} electric fields
${\bm E}(t)$ that oscillate around zero, 
whereas Eq.~(\ref{HESRmuBoft}) holds also for static $B$-fields.
The reason why a static electric field ${\bm E}$ cannot be used in Eq.~(\ref{HEDSRnuEoft})
will become clear after we explain the origin of $\mbox{\boldmath $\nu$}$ in Eq.~(\ref{HEDSRnuEoft}).

The spin-electric moment $\mbox{\boldmath $\nu$}$ arises because the dipolar
transitions in the quantum dot become allowed, e.g., for the ground state
\begin{equation}
\langle\psi_{0\uparrow}|{\bm r}|\psi_{0\downarrow}\rangle\neq 0.
\end{equation}
The electron charge density operator $\rho ({\bm r})=-e\delta({\bm r}-{\bm r}_{el})$,
where ${\bm r}_{el}$ is the electron coordinate,
acquires spin-dependent terms in the transformed basis,
\begin{eqnarray}
&&|\psi_{ns}\rangle=e^{-S}|\psi_n\rangle|\chi_s\rangle,\\
&&\tilde \rho ({\bm r})=e^{S}\rho ({\bm r})e^{-S},
\end{eqnarray}
where $e^{-S}$ is the transformation used in Section~\ref{secSpElCo} and studied in
detail in Appendix~\ref{appSW}.
One can present $\tilde \rho ({\bm r})$ as a sum of two terms,
\begin{equation}
\tilde \rho ({\bm r})=\bar\rho ({\bm r})+\delta\rho ({\bm r}),
\end{equation}
where $\bar\rho ({\bm r})$ is spin independent and $\delta\rho ({\bm r})$ is proportional to the spin.
Then the spin-electric moment can be written as follows,
\begin{equation}
{\bm \nu}=\int {\bm r}\delta\rho ({\bm r})dv,
\label{defofnusigmasigma}
\end{equation}
where $dv$ is the elementary volume of integration.
Equation (\ref{defofnusigmasigma}) unveils the physical meaning
of the spin-electric moment ${\bm\nu}$:
due to the mixed spin and orbit nature of the electron density,
the electron spin couples to the first moment (dipole moment) of the electron.

While oscillating around an equilibrium position, the electron produces a time-dependent
dipole moment, part of which is proportional to the electron spin.
Obviously, in a static electric field, one can set to zero the electron dipole moment, because the
new electron position can be taken as the equilibrium one. 
Therefore, only the change of the moment as a function of time has a physical meaning for single
electrons (as for any other monopoles).
In contrast, the electron magnetic moment ${\bm \mu}$ couples to static magnetic fields, because
one can view ${\bm \mu}$ as arising from a pair of Dirac monopoles of opposite signs, for
which the relative distance between them has an absolute meaning.

Equation (\ref{defofnusigmasigma}) is written in a very general (operator) form.
After taking the expectation value in the orbital ground state $|\psi_0\rangle$,
we obtain
\begin{eqnarray}\label{mbnum12barbarnusigma}
&&{\bm \nu}=-\frac{1}{2}\bar{\bar{\nu}}\mbox{\boldmath $\sigma$},\\
&&\bar{\bar{\nu}}_{ij}=2e\frac{\partial}{\partial\sigma_j}
\langle\psi_0|e^{S}r_ie^{-S}|\psi_0\rangle,
\end{eqnarray}
where the derivative with respect to $\sigma_j$ is defined as
a usual derivative of an expression that is linear in ${\bm \sigma}$.
Note that Eq.~(\ref{mbnum12barbarnusigma}) is analogous to Eq.~(\ref{defofmugmuBsigma})
where the role of $g\mu_B$ is played by the tensor $\bar{\bar{\nu}}$.
Using Eqs.~(\ref{Heffexpl})-(\ref{Omegaexpl}) we obtain for the linear in momentum spin-orbit
interaction,
\begin{equation}
\bar{\bar{\nu}}_{ij}=-\frac{2eg\mu_B}{m_e\omega_0^2}\varepsilon_{jkl}B_k{\left(\lambda_{SO}^{-1}\right)}_{li}.
\end{equation}
Similarly, for the $p^3$ Dresselhaus terms we obtain from
Eq.~(\ref{eq12htseqEE})
\begin{equation}
\bar{\bar{\nu}}_{ij}=\frac{2e\gamma\eta\hbar^2\omega_c}{9m_e\omega_0^4}
\left(\begin{array}{cc}
0&1\\
1&0
\end{array}\right),
\end{equation}
where we use the coordinate frame $(x,y)$ to represent the tensor.
Note that in both cases $\bar{\bar{\nu}}_{ij}$ is proportional to
the magnetic field (or one of its components).
The spin-orbit interaction produces no spin-electric coupling at $B=0$,
because of the time-reversal symmetry of spin-orbit interaction.

The analogy between ${\bm \mu}$ and ${\bm \nu}$
is also seen in the pairwise interaction between spins in
separate (not tunnel coupled) quantum dots.~\cite{SpinSpin}
For an unscreened Coulomb interaction between electrons,
the spin-spin interaction is analogous to the
magnetic dipole-dipole interaction,~\cite{SpinSpin}
\begin{equation}
H_{\rm dd}=\sum_{i<j}\frac{{\bm \nu}_i\cdot{\bm \nu}_jr_{ij}^2
-3({\bm \nu}_i\cdot{\bm r}_{ij})({\bm \nu}_j\cdot{\bm r}_{ij})}
{\kappa r_{ij}^5},
\label{SpinSpinDipolar}
\end{equation}
where ${\bm r}_{ij}={\bm r}_i-{\bm r}_j$ is the distance between two
quantum dots ($r_{ij}\gg\lambda_d$) and $\kappa$ is the electric permittivity of the material.
For further detail and a microscopic derivation of
Eq.~(\ref{SpinSpinDipolar}) we refer the reader to Ref.~\onlinecite{SpinSpin}.

Next we discuss the limitations of our theory.
Throughout the paper, we have assumed that the spin
orbit interaction is weak compared to the dot level spacing,
or, in other words, that $\lambda_d/\lambda_{SO}\ll 1$.
This assumption allowed us to use the perturbation theory
to find the unitary transformation $\exp(-S)$, 
see Appendix~\ref{appSW}. 
Of course, this restriction was not necessary, if, e.g., we were to
apply numerical methods to diagonalize the Hamiltonian.
In particular,
note that Eq.~(\ref{defofnusigmasigma}) is meaningful
whenever the unitary transformation $\exp(-S)$ exists. 
The latter is always the case, including also extended states.
Our perturbative results are qualitatively correct for $\lambda_d\lesssim \lambda_{SO}$
and can be used in experiments to estimate the strength of the EDSR effect.
The case $\lambda_d\gg \lambda_{SO}$ is more seldom and requires a separate
theoretical investigation.

As a second limitation, we would like to mention the adiabaticity criterion.
In Sections~\ref{secSpElCo} and~\ref{secSOp3},
we have derived effective Hamiltonians for the low energy subspace of the quantum dot
Hilbert space.
For the validity of this effective description, it is important that 
the switching (on and off) of the effective interaction occurs 
on a time scale that is larger than the inverse level spacing in the quantum dot.
Obviously, this criterion excludes applicability of our theory to extended states.
In practice, however, the finite temperature $T$ imposes a more stringent criterion 
on the confinement energy, $\hbar^2/m_e\lambda_d^2\gg k_{B}T$.

The third limitation of our theory is a small amplitude of oscillation of the quantum dot,
$r_0/\lambda_{SO}\ll 1$. 
We have shown in Section~\ref{secSpElCo} that the EDSR effect
is proportional to this small parameter.
Thus, for breaking the time-reversal symmetry by the Zeeman interaction we have
(by order of magnitude)
\begin{equation}
\omega_R\sim \omega_Z\frac{r_0}{\lambda_{SO}},
\label{prelast}
\end{equation}
where $\omega_Z=E_Z/\hbar$.
Similarly, for breaking the time-reversal symmetry by the orbital $B$-field effect (Section~\ref{secSOp3}),
we have
\begin{equation}
\omega_R\sim \omega_c\frac{r_0}{\lambda_{SO}}\left\langle R/r\right\rangle,
\label{last}
\end{equation}
where $\left\langle R/r\right\rangle$ 
is the small parameter of deviation of the quantum dot confinement from harmonic.
We remark that our theory remains qualitatively valid also for $r_0/\lambda_{SO}\sim 1$
and for $\left\langle R/r\right\rangle\sim 1$. 
Beyond these limits, we do not expect the Rabi frequency to grow indefinitely.
The Rabi frequency is bound in the case of Eq.~(\ref{prelast}) by $\omega_R\leq\omega_Z$,
and in the case of Eq.~(\ref{last}) by $\omega_R\leq\omega_c$.
We conclude that, by designing quantum dot setups that allow for large 
oscillation amplitudes $r_0\lesssim\lambda_{SO}$,
the EDSR effect can be strongly enhanced, beyond the numeric estimates made in 
Sections~\ref{secSpElCo}, \ref{secDynDec}, and~\ref{secSOp3}.

In conclusion, the EDSR mechanisms presented above provides 
a means of implementing local electrical control of electron spins in quantum dots.

\section*{Acknowledgment}
We thank M. Trif, L.M.K. Vandersypen, and D.M. Zumb\"{u}hl for discussions.
We acknowledge support from the Swiss NSF, NCCR Nanoscience, DARPA, ONR, and JST ICORP.

\appendix
\section{Schrieffer-Wolff transformation and fine structure
}                      %
\label{appSW}
In this Appendix, we first work out the Schrieffer-Wolff transformation 
to the third order of perturbation theory and for a general weak perturbation.
Then, we consider an example Hamiltonian and use the Schrieffer-Wolff 
transformation to partly diagonalize the Hamiltonian.
Finally, we analyze the fine structure of the transformed Hamiltonian
and complete its diagonalization by an additional unitary transformation.

As in standard perturbation theory, we consider a Hamiltonian $H=H_0+H_1$, 
where $H_1$ is a weak perturbation with respect to $H_0$.
For the matrix elements of $H_1$, we assume
\begin{eqnarray}\label{H1weak0}
&&\langle n|H_1|m\rangle=0,\;\;\; \mbox{for}\;\;\; E_n=E_m,\\
&&\langle n|H_1|m\rangle\ll E_n- E_m,\;\;\; \mbox{for}\;\;\; E_n\neq E_m,
\label{H1weakneq}
\end{eqnarray}
where $|n\rangle$ and $E_n$ are, respectively, the eigenstates and eigenvalues of
$H_0$, and are obtained from $H_0|n\rangle=E_n|n\rangle$.

The projector ${\cal P}$, defined as follows
\begin{equation}
{\cal P}A=\sum_{\genfrac{}{}{0pt}{3}{nm}{E_n=E_m}}A_{nm}|n\rangle\langle m|,\;\;\;\;\;\;\;\;\; \forall A
\label{defPcalA}
\end{equation}
projects onto the diagonal or degenerate part of $H_0$.
In the particular case, when the spectrum of $H_0$ is non-degenerate,
${\cal P}$ assumes ${\cal P}A=\sum_nA_{nn}|n\rangle\langle n|$, $\forall A$.
From Eq.~(\ref{H1weak0}) and the definition (\ref{defPcalA}), it follows that
\begin{eqnarray}
{\cal P}H_1&=&0,\\
{\cal P}H_0&=&H_0.
\end{eqnarray}

Next we look for a unitary transformation that
brings the Hamiltonian $H=H_0+H_1$ to a partly diagonal form,
\begin{equation}\label{AppSWtildeHH}
\tilde H=e^{S'}\left(H_0+H_1\right)e^{-S'}=H_0+\Delta H,
\end{equation}
where the operator $\Delta H$ obeys ${\cal P}\Delta H=\Delta H$.
Here, $S'=-S'^\dagger$ is the transformation matrix.
The unitary transformation in Eq.~(\ref{AppSWtildeHH}) is
called the Schrieffer-Wolff transformation.~\cite{SW}
We expand $S'$ and $\Delta H$ in terms of the perturbation $H_1$:
\begin{eqnarray}\label{APPSWSS123}
&&S'=S'^{(1)}+S'^{(2)}+S'^{(3)}+\dots,\\
&&\Delta H=\Delta H^{(1)}+\Delta H^{(2)}+\Delta H^{(3)}+\dots,
\label{APPSWDH123}
\end{eqnarray}
where the superscripts give the order of perturbation theory.
Substituting Eqs.~(\ref{APPSWSS123}) and (\ref{APPSWDH123}) into Eq.~(\ref{AppSWtildeHH}),
we find a set of equations for $S'$,
\begin{eqnarray}\label{APPSWeqS1}
[H_0,S'^{(1)}]&=&H_1,\\
{}[H_0,S'^{(2)}]&=&\frac{{\cal Q}}{2}[S'^{(1)},H_1],\\
\label{APPSWeqS2}
{}[H_0,S'^{(3)}]&=&\frac{{\cal Q}}{2}[S'^{(2)},H_1]+\frac{{\cal Q}}{12}[S'^{(1)},[S'^{(1)},H_1]]+\nonumber\\
&&\frac{{\cal Q}}{4}[S'^{(1)},{\cal P}[S'^{(1)},H_1]],\label{APPSWeqS3}
\end{eqnarray}
where ${\cal Q}\equiv 1-{\cal P}$.
It is important to note that $S$ is defined in Eqs.~(\ref{APPSWeqS1})-(\ref{APPSWeqS3}) 
up to terms ${\cal P}M$, where $M$ is arbitrary.
Such terms drop out on the left-hand side in Eqs.~(\ref{APPSWeqS1})-(\ref{APPSWeqS3})
because $[H_0,{\cal P}S']=0$.
Thus, ${\cal P}S'$ can be chosen arbitrarily, which shows that 
there are infinitely many transformation matrices $S'$ that satisfy Eq.~(\ref{AppSWtildeHH}).
For simplicity, we choose ${\cal P}S'=0$ and address the fine structure of
$\tilde{H}=H_0+\Delta H$ later on.
For the operator $\Delta H$, we obtain
\begin{eqnarray}
\Delta H^{(1)}&=&0,\\
\Delta H^{(2)}&=&\frac{{\cal P}}{2}[S'^{(1)},H_1],\\
\Delta H^{(3)}&=&\frac{{\cal P}}{3}[S'^{(1)},[S'^{(1)},H_1]].
\end{eqnarray}

Introducing the Liouvillean $\hat{L}_0$: $\hat{L}_0A=[H_0,A]$, $\forall A$,
we can formally solve Eqs.~(\ref{APPSWeqS1})-(\ref{APPSWeqS3}) one by one.
For example, the transformation matrix at the lowest order reads
$S'^{(1)}={\cal Q}\hat{L}_0^{-1}H_1$.
For $\Delta H$, we recover then the perturbation theory expansion in a more familiar form,
\begin{equation}
\Delta H=-{\cal P}H_1\hat{L}_0^{-1}H_1+{\cal P}H_1\hat{L}_0^{-1}
H_1\hat{L}_0^{-1}H_1+\dots,
\label{DHperttheorypm}
\end{equation}
with the usual convention, ${\cal P}\hat{L}_0^{-1}A=0$, $\forall A$, adopted.

Next, we remark that the fine structure of $\tilde{H}=H_0+\Delta H$ can be addressed
in each particular case by means of degenerate perturbation theory.
As an example, we consider here the Hamiltonian $H=H_0+H_1$, 
with $H_0=H_d+H_Z$ and $H_1=H_{SO}$.
Here, $H_d$ is given in Eq.~(\ref{Hdot}), with $U({\bm r})=m_e\omega_0^2r^2/2$,
$H_Z$ is given in Eq.~(\ref{HZ}), and $H_{SO}$ is given in Eq.~(\ref{HSO}).
Using the transformation matrix $S'=S$, with $S$ given in Eq.~(\ref{Smatrixxizeta}),
we obtain a diagonal Hamiltonian, $\tilde{H}=H_d+H_Z$, 
at the first order of $H_{SO}$.
At the second order of $H_{SO}$, however, a fine structure in the energy
spectrum arises.
At $B=0$, the transformed Hamiltonian reads
\begin{equation}
\tilde H=\frac{p^2}{2m_e}+\frac{m_e\omega_0^2}{2}r^2+\frac{1}{2}\Delta_{SO}\ell_z\sigma_z,
\label{eqtHDSOfinestruc}
\end{equation}
where $\ell_z=-i(x\partial/\partial y-y\partial/\partial x)$ is the electron rotational momentum and
$\Delta_{SO}=2m_e(\beta^2-\alpha^2)$.
The Kramers doublets are identified, in this case, as the pairs of states with 
quantum numbers $(\ell_z,\sigma_z)$ and $(-\ell_z,-\sigma_z)$.
For $\ell_z>0$, the two-fold orbital degeneracy is lifted and a
splitting $\Delta_{SO}\ell_z$ arises.
Note that the ground orbital state, which has $\ell_z=0$, remains
doubly degenerate in this case.

At $B\neq 0$, the fine-structure interaction in Eq.~(\ref{eqtHDSOfinestruc}) 
is modified by both the Zeeman energy $E_Z$ and the cyclotron frequency $\omega_c$.
For simplicity, we omit terms $\sim\Delta_{SO}E_Z/\hbar\omega_0$, but keep terms 
$\sim\Delta_{SO}\omega_c/\omega_0$, assuming that $E_Z\ll\hbar\omega_c$.
Then, the Hamiltonian (\ref{eqtHDSOfinestruc}) acquires two extra terms
\begin{equation}
\frac{E_Z}{2}\mbox{\boldmath $n$}\cdot \mbox{\boldmath $\sigma$}
+\frac{\Delta_{SO}}{4\lambda^2}\sigma_z{\cal P}r^2,
\label{eqtHDSOfinestruc2}
\end{equation}
where $\lambda=\sqrt{\hbar/m_e\omega_c}$ is the magnetic length and
we use the symmetric gauge, ${\bm A}({\bm r})=B_z(-y/2,x/2,0)$.
The last term in Eq.~(\ref{eqtHDSOfinestruc2}) can be viewed as a 
renormalization of the electron $g$-factor.
Allowing for an anisotropic Zeeman interaction,
\begin{equation}
H_Z^{\rm eff}=\frac{1}{2}\mu_B\sum_{ij}g_{ij}\sigma_iB_j,
\label{renormZeemangijsB}
\end{equation}
we obtain that the tensor $g_{ij}$ is diagonal in the
main crystallographic frame, with
\begin{eqnarray}
&&g_{xx}=g_{yy}=g,\nonumber\\
&&g_{zz}=g+\frac{m\Delta_{SO}}{\hbar^2}\langle \psi_n|r^2|\psi_n\rangle,
\label{gijtensorcorr}
\end{eqnarray}
where $m$ is the electron mass in vacuum and $\psi_n$ is the electron
orbital state.
For the ground orbital state, the corrected $g$-factor reads
$g_{zz}=g+m\Delta_{SO}/m_e\hbar\omega$, where $\omega=\sqrt{\omega_0^2+\omega_c^2/4}$.
Note that the sign of the correction is given by the sign of
$\beta^2-\alpha^2$ contained in $\Delta_{SO}$.
The spin quantization axis does not, in general, coincide with the magnetic field
direction ${\bm n}$ and is given by the following unit vector
\begin{equation}
\tilde{\bm n}=\frac{{\bm n}+\zeta{\bm n}_z}{\sqrt{1+\zeta(2+\zeta)n_z^2}},
\end{equation}
where $\zeta=(g_{zz}-g)/g$.
An additional unitary transformation can be used to diagonalize
the $2\times 2$ blocks of Zeeman-split Kramers doublets,
\begin{equation}
\tilde H_Z^{\rm eff}=e^{S\mbox{\scriptsize $''$}}H_Z^{\rm eff}e^{-S\mbox{\scriptsize $''$}}=\frac{1}{2}\tilde E_Z
{\bm n}\cdot\mbox{\boldmath $\sigma$}
\end{equation}
where $\tilde E_Z=E_Z\sqrt{1+\zeta(2+\zeta)n_z^2}$ is the renormalized Zeeman energy and 
\begin{equation}
e^{-S\mbox{\scriptsize $''$}}=\sqrt{\frac{1+{\bm n}\cdot\tilde{\bm n}}{2}}-i\frac{\zeta[{\bm n}\times{\bm n}_z]\cdot\mbox{\boldmath $\sigma$}}
{\sqrt{\zeta^2n_z^2(1-n_z^2)}}\sqrt{\frac{1-{\bm n}\cdot\tilde{\bm n}}{2}}.
\end{equation}
So far, we have considered a given orbital state $\psi_n$, for which the tensor
$g_{ij}$ is given in Eq.~(\ref{gijtensorcorr}).
The transformation above is also valid in general, provided $\zeta$ is
understood as a diagonal operator, $\zeta=(m\Delta_{SO}/g\hbar^2){\cal P}r^2$.

We summarize by mentioning that the unitary transformation in Eq.~(\ref{AppSWtildeHH})
can, in principle, be adjusted to give a fully diagonal $\tilde H=H_0+\Delta H$, 
i.e. we had not to require ${\cal P}S'=0$ in the first place.
However, in practice, it is more convenient first to apply the non-degenerate
perturbation theory, Eqs.~(\ref{APPSWeqS1})-(\ref{DHperttheorypm}), and
then, at the end, complete the diagonalization of $H_0+\Delta H$
by a second unitary transformation.
The latter is specific to each particular case and amounts, in general, to 
applying the degenerate perturbation theory.
For the sake of simplicity, we shall refer to $S$ in the main text of the paper 
as to the full transformation matrix, despite the fact that the respective
unitary transformation comes, in practice, as a product of two unitary transformations.
Thus, we denote the product $e^{-S'}e^{-S''}$ by $e^{-S}$ in the main text.
Finally, we remark that $e^{-S''}\approx 1+{\cal O}(H_{SO}^2)$.


\begin{references}
\bibitem{Wolf}
S.A. Wolf, D.D. Awschalom, R.A. Buhrman, J.M. Daughton, S. von Moln\'{a}r, M.L. Roukes, A.Y. Chtchelkanova, and D.M. Treger, 
Science {\bf 294}, 1488 (2001).

\bibitem{ALS}
{\it Semiconductor Spintronics and Quantum Computation},
D.D. Awschalom, D. Loss, and N. Samarth (eds.), (Springer, Berlin, 2002).

\bibitem{LD}
D. Loss and D.P. DiVincenzo, Phys. Rev. A \(\textbf{{57}}\),
120 (1998).

\bibitem{Atherton}
N.M.~Atherton, {\em Electron Spin Resonance},
(Ellis Horwod Limited, New York, 1973).

\bibitem{Engel}
H.-A. Engel and D. Loss, 
Phys. Rev. Lett. {\bf 86}, 4648 (2001);
Phys. Rev. B {\bf 65}, 195321 (2002).

\bibitem{Kato}
Y.~Kato, R.C.~Myers, D.C.~Driscoll, A.C.~Gossard, J.~Levy, and D.D.~Awschalom,
Science {\bf 299}, 1201 (2003).

\bibitem{Miller}
R.C. Miller, A.C. Gossard, D.A. Kleinman, and O. Munteanu 
Phys. Rev. B {\bf 29}, 3740 (1984).

\bibitem{Salis}
G. Salis, Y. Kato, K. Ensslin, D.C. Driscoll, A.C. Gossard, and D.D. Awschalom,
Nature (London) {\bf 414}, 619 (2001).

\bibitem{RashbaEfros}
E.I.~Rashba and Al.L.~Efros,
Phys. Rev. Lett. {\bf 91}, 126405 (2003);
Appl. Phys. Lett. {\bf 83}, 5295 (2003);
E.I.~Rashba, 
J. Supercond. Incorp. Novel. Magn. {\bf 18}, 137, (2005).

\bibitem{Dress}
G. Dresselhaus, 
Phys. Rev. {\bf 100}, 580 (1955).

\bibitem{Rashba}
Y. Bychkov and E. I. Rashba, 
J. Phys. C {\bf 17}, 6039 (1984).

\bibitem{Bell}
R.L.~Bell,
Phys. Rev. Lett. {\bf 9}, 52 (1962).

\bibitem{McCombe}
B.D.~McCombe, S.G.~Bishop, and R.~Kaplan,
Phys. Rev. Lett. {\bf 18}, 748 (1967).

\bibitem{Dobrowolska}
M.~Dobrowolska, H.D.~Drew, J.K.~Furdyna, T.~Ichiguchi, A.~Witowski, and P.A.~Wolff, 
Phys. Rev. Lett. {\bf 49}, 845 (1982);
M.~Dobrowolska, A.~Witowski, J.K.~Furdyna, T.~Ichiguchi, H.D.~Drew, and P.A.~Wolff, 
Phys. Rev. B {\bf 29}, 6652 (1984).

\bibitem{Schulte}
M.~Schulte, J.G.S.~Lok, G.~Denninger, and W.~Dietsche,
Phys. Rev. Lett. {\bf 94}, 137601 (2005).

\bibitem{DuckheimLossEDSR}
M.~Duckheim and D.~Loss,
Nature Physics {\bf 2}, 195 (2006).

\bibitem{KatoEDSR}
Y.~Kato, R.C.~Myers, A.C.~Gossard, and D.D.~Awschalom,
Nature (London) {\bf 427}, 50 (2004).

\bibitem{Nitta}
J. Nitta, T. Akazaki, H. Takayanagi, and T. Enoki
Phys. Rev. Lett. {\bf 78}, 1335 (1997).

\bibitem{KAT}
L.P. Kouwenhoven, D.G. Austing, and S. Tarucha,
Rep. Prog. Phys. {\bf 64} 701 (2001).

\bibitem{KhNaHaAlFa}
A.V. Khaetskii and Yu.V. Nazarov,
Phys. Rev. B {\bf 61}, 12639 (2000);
B.I. Halperin, A. Stern, Y. Oreg, J.N.H.J. Cremers, J.A. Folk, and C.M. Marcus,
Phys.\ Rev.\ Lett.\ {\bf 86}, 2106 (2001);
I.L. Aleiner and V.I. Fal'ko,  Phys.\ Rev.\ Lett.\ {\bf 87}, 256801 (2001).

\bibitem{GKL}
V.N. Golovach, A. Khaetskii, and D. Loss, 
Phys. Rev. Lett. {\bf 93}, 016601 (2004).

\bibitem{KhaNaz}
A.V. Khaetskii and Yu.V. Nazarov,
Phys. Rev. B {\bf 64}, 125316 (2001).

\bibitem{DyakKach}
M.I. D'yakonov and V.Yu. Kachorovskii,
Sov. Phys. Semicond. {\bf 20}, 110 (1986).

\bibitem{ENature}
J.M. Elzerman, R. Hanson, L.H. Willems van Beveren, B. Witkamp, L.M.K. Vandersypen, and L.P. Kouwenhoven,
Nature (London) {\bf 430}, 431 (2004).

\bibitem{Kroutvar}
M. Kroutvar, Y. Ducommun, D. Heiss, M. Bichler, D. Schuh, G. Abstreiter, and J.J. Finley,
Nature (London) {\bf 432}, 81 (2004).

\bibitem{Jackson}
J.D. Jackson, {\em Classical Electrodynamics},
(John Wiley \& Sons, Inc., New York, 1999).

\bibitem{noteDelft}
F.H.L.~Koppens, C.~Buizert, K.J.~Tielrooij, I.T.~Vink, K.C.~Nowack, T.~Meunier, L.P.~Kouwenhoven, and L.M.K.~Vandersypen,
Nature (London) \textbf{442}, 766 (2006).

\bibitem{BGL}
M.~Borhani, V.N.~Golovach, and D.~Loss,
Phys. Rev. B {\bf 73}, 155311 (2006).

\bibitem{Bloch}
F.~Bloch, Phys. Rev. {\bf 70}, 460 (1946);
R. K. Wangsness and F. Bloch, Phys. Rev. {\bf 89}, 728 (1953);
F.~Bloch, Phys. Rev. {\bf 105}, 1206 (1957).

\bibitem{CohenTannoudji}
C.~Cohen-Tannoudji, B.~Diu, and F.~Lalo\"e, {\em Quantum Mechanics}, 
Vol. II (John Wiley \& Sons, New York, 1977).

\bibitem {AmashaZumbuhl}
S.~Amasha, K.~MacLean, I.~Radu, D.M.~Zumbuhl, M.A.~Kastner, M.P.~Hanson, and A.C.~Gossard,
cond-mat/0607110.

\bibitem{PikusTitkov}
G.E.~Pikus and A.N.~Titkov,
in {\em Optical orientation}, edited by F. Meier and B.P. Zakharchenya 
(North-Holland, Amsterdam, 1984).

\bibitem{DefQuant}
F.~Bayen, M.~Flato, C.~Fronsdal, A.~Lichnerowicz, and D.~Sternheimer,
Ann. Phys. {\bf 111}, 61 (1978);
Ann. Phys. {\bf 111}, 111 (1978).

\bibitem{SpinSpin}
M.~Trif, V.N.~Golovach, and D.~Loss,
cond-mat/0608512.

\bibitem{SW}
J.R.~Schrieffer, P.A.~Wolff, Phys. Rev. {\bf 149}, 491 (1966).

\end{references}
\end{document}